\documentclass[aps, 
			twocolumn,
			noeprint
			amssymb, 
			]{revtex4-1}

\usepackage[version=3]{mhchem} 
\usepackage[T1]{fontenc} 
\usepackage{graphicx}
\usepackage{gensymb}
\usepackage{xfrac}
\usepackage{xcolor}
\usepackage{multirow}

\begin{document}

\title{A New Family of Two-Dimensional Crystals: Open-Framework T$_{3}$X (T=C, Si, Ge, Sn; X=O, S, Se, Te) Compounds with Tetrahedral Bonding}

\author{Kisung Chae}
\affiliation{Korea Institute for Advanced Study, Seoul 02455, South Korea}

\author{Young-Woo Son}
\email{hand@kias.re.kr}
\affiliation{Korea Institute for Advanced Study, Seoul 02455, South Korea}


\date{\today}

\begin{abstract}
To accelerate development of innovative materials, their modelings and predictions with useful functionalities are of vital importance. 
Here, based on a recently developed crystal structure prediction method, we find a new family of stable two-dimensional crystals with an open-channel tetrahedral bonding network, rendering a potential for electronic and energy applications. 
The proposed structural prototype with a space group of Cmme hosts at least thirteen different freestanding T$_{3}$X compounds with group IV (T=C, Si, Ge, Sn) and VI (X=O, S, Se, Te) elements. 
Moreover, the proposed materials display diverse electronic properties ranging from direct band gap semiconductor to topological insulator at their pristine forms, which are further tunable by mechanical strain. 
\end{abstract}

\maketitle

\section{Introduction}
Two-dimensional (2D) materials with a-few-atom thickness show interesting physical properties and offer great potentials for various applications. 
After successfully isolating a single layer from several layered materials such as graphene~\cite{Novoselov10451} and molybdenum disulfide~\cite{C6NR02253G}, tremendous efforts have been devoted to discover other 2D materials, expecting new functionalities and enhanced performance for new materials and their heterostructures. 
Alongside experimental progress in isolating new 2D crystals, materials discovery by means of high-throughput computations or predictive computational methods has also been actively pursued to find new materials~\cite{2053-1583-5-4-042002, doi:10.1021/acs.nanolett.6b05229, PhysRevLett.118.106101, Mounet2018}.

Despite those recent efforts, however, database size of the 2D materials is still limited compared to that of the three-dimensional (3D) materials.
In addition, significant portion of the 2D materials registered in the materials data repositories are categorized by specific prototypes such as iron oxychloride (a space group of Pmmn), molybdenum disulfide (P$\bar{3}$m2) and cadmium iodide (P$\bar{3}$m1).
For instance, one 2D materials database (C2DB - http://c2db.fysik.dtu.dk) shows that the above three categories take up over half of the total items registered, and they reach even to $\sim$60\% among the stable ones~\cite{2053-1583-5-4-042002}.
On one hand, generation of the database could be biased because the data were generated by decorating known prototypical structures with stoichiometric elements, indicating that chemical space was not uniformly sampled.
On the other hand, it also means that crystal structure prediction (CSP), offering new 2D prototypes for a number of novel materials, can play an important role to expand both material and property spaces.

Global optimization such as evolutionary algorithm~\cite{JM9950501269} and particle swarm optimization~\cite{PhysRevLett.57.2607} is a useful method to predict new crystal structures, making advances to expand 3D material space~\cite{PhysRevB.82.094116, doi:10.1063/1.2210932}.
Even though it works fairly well with condensed bulk materials (e.g., high-pressure phases), low-dimensional materials with reduced densities such as porous media and layered materials are difficult to be found due to the enormously expanded search space.
This can be understood by an analogy of CSP to the number of combinatorial ways to arrange $m$ identical atoms in a fixed-volume unit cell containing the $n$ grids~\cite{doi:10.1063/1.2210932}, i.e., $\dfrac{n!}{m!(n-m)!}$.
Increasing $n$ for a given $m$, which is the case for low-dimensional materials, exponentially increases the number of possible configurations.

Recently, we developed a new CSP method particularly suited for 2D systems~\cite{2053-1583-5-2-025013}: Search by \emph{Ab initio} Novel Design \emph{via} Wyckoff positions Iteration in Conformational Hypersurface (SANDWICH).
In this new method, the 2D crystals are constructed by joining both top and bottom surface layers while the space between the surface layers is filled with bridge atoms~\cite{2053-1583-5-2-025013}.
The inner part is constructed by exhaustive enumeration of a subset of the Wyckoff positions corresponding to the space group determined by the surface layers.
Our method has a merit over the conventional method in finding new 2D crystals with a stable open-framework of crystal lattice since the global optimization with a single element hardly find a stable hollow structure without \emph{a priori} inputs from experiments~\cite{PhysRevB.95.155426}.
Using the new method, we could generate a series of novel stable 2D silicon crystals with a large hollow channel as well as with an atomically flat surface showing interesting electronic structures~\cite{2053-1583-5-2-025013}.

In this Letter, we find theoretically a prototypical 2D crystal family which can host various T$_{3}$X compounds, where T and X stand for group IV (C, Si, Ge, Sn) and group VI (O, S, Se, Te) elements, respectively. 
The newly found 2D materials show an atomic arrangement with a space group of Cmme (No. 67), which is very distinctive compared with any other known 2D materials, and features an open-framework owing to strong covalent bonding characters of group IV elements. 
We confirmed that most of the monolayer T$_{3}$Xs are dynamically stable by using phonon dispersion relations, and showed their structural robustness more rigorously with \emph{ab initio} molecular dynamics (AIMD) simulations.
We also find not only that they can form stable layered bulk phases, but also that their exfoliation energies are moderate enough, so that single layers can be easily isolated from their bulk materials.
In particular, the electronic properties vary significantly in a wide range, depending on their chemical compositions, which are further shown to be tunable by mechanical strain.

\section{Computational Details}
We found a new family of T$_{3}$X using our approach that was previously used to predict a series of 2D Si crystals. 
It was shown that one of the new 2D Si crystals with oxidized surfaces is extraordinary stable and has a direct band gap~\cite{2053-1583-5-2-025013}.
By using the oxidized structure as a prototype, we generated other 2D materials by substituting Si and O with other elements having the same valence electron configurations: group IV (carbon group or tetragen) and group VI (oxygen group or chalcogen), respectively. 
Following a nomenclature for other 2D systems such as graphene, phosphorene and MXene, we name the new series of 2D tetragen-chalcogen T$_{3}$X compounds as TXene. 

The generated structures of TXenes were optimized by conjugate gradient method until the Hellmann--Feynman force on each of the atoms becomes smaller than 1 meV/\AA.
Total energy and forces were obtained by performing density functional theory calculations as implemented in Vienna \emph{Ab initio} Software Package~\cite{PhysRevB.54.11169, KRESSE199615} code.
For structural optimization, Perdew-Burke-Ernzerhof (PBE) exchange-correlation functional~\cite{PhysRevLett.77.3865} was used with rVV10 nonlocal functional correction when necessary~\cite{PhysRevB.87.041108}, while hybrid functional (HSE06)~\cite{doi:10.1063/1.2404663} was used for computing electronic structures.
Spin-orbit interaction was switched on for systems with inverted bands at Fermi level.
Kinetic energy cutoff of 500 eV was used for plane-wave expansion, and projector augmented wave method was used for atom cores~\cite{PhysRevB.59.1758}. 
The \emph{k}-points in reciprocal space were uniformly sampled including the $\Gamma$-point with their spacing less than or equal to 0.05 \AA$^{-1}$.
Electron density for a given system was regarded to be self-consistent if change of both each eigenvalue and total energy compared to the previous step is smaller than 10$^{-8}$ eV. 

To confirm dynamical stability, we also computed harmonic phonon dispersions of T$_{3}$X by using frozen phonon method~\cite{PhysRevLett.78.4063} as implemented in phonopy~\cite{TOGO20151} code. 
Supercells with a size of 3$\times$3$\times$1 were used to compute phonon dispersions and AIMD simulations.
We used Nose-Hoover thermostat with integration timestep of 1 femto second.
Exfoliation energy ($E_{\mathrm{exf}}$) with nonlocal correction of rVV10~\cite{PhysRevB.87.041108} was computed using a newly developed method suggested in Ref.~\cite{doi:10.1021/acs.nanolett.7b04201}
\begin{equation}
E_{\mathrm{exf}} = \dfrac{E_{\mathrm{mono}}-E_{\mathrm{bulk}}}{A_{\mathrm{bulk}}},
\label{eq1}
\end{equation}
where E and A refer to total energy and area of the unitcell, respectively.

\section{Results and Discussion}
\begin{figure}[]
	\centering
	\includegraphics[width=\columnwidth]{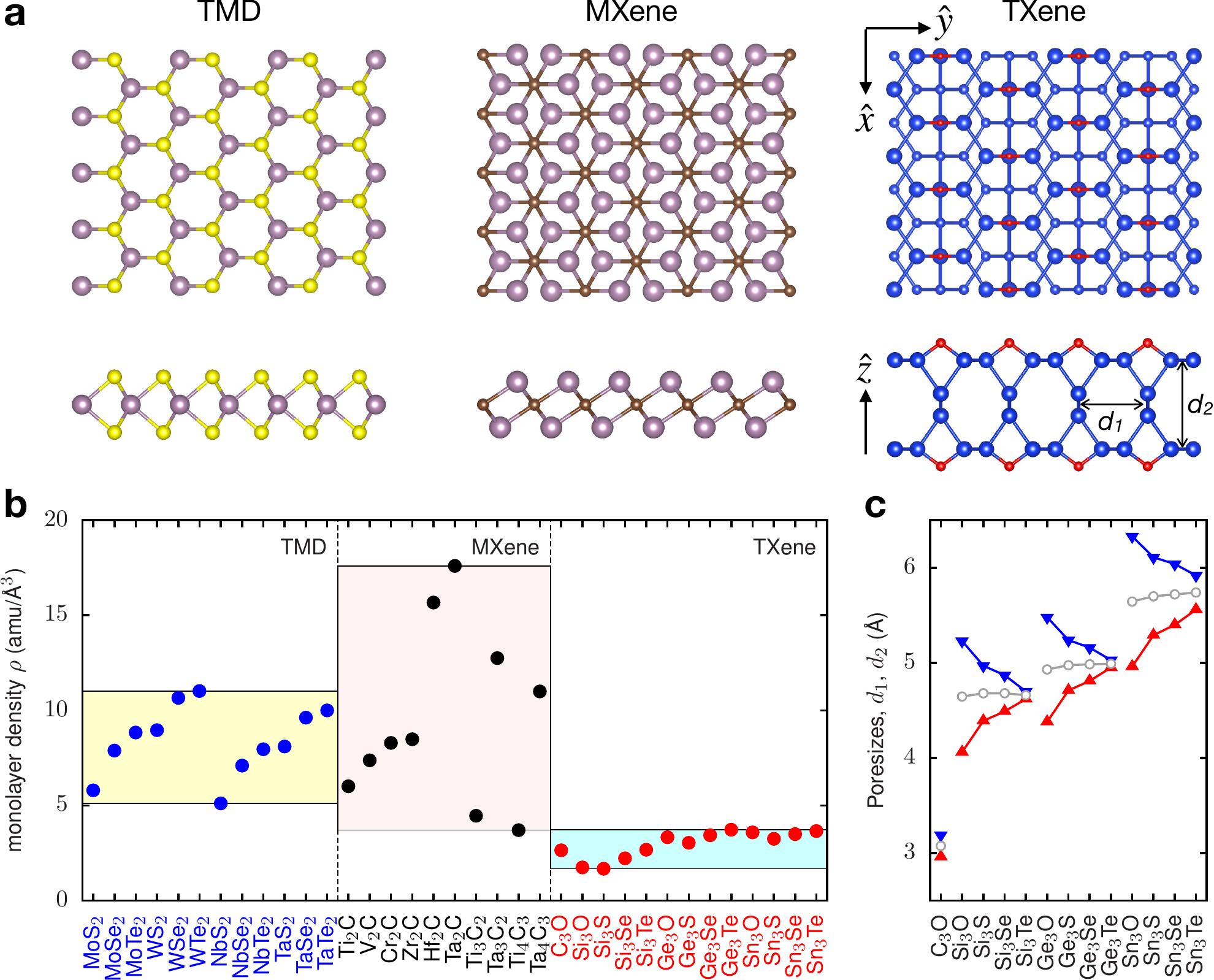}
	\caption{Structural features: (a) atomic arrangements of transition metal dichalcogenide (TMD), transition metal carbonitride (MXene) and tetragen chalcogenide (TXene). For TXene, size of the atoms were modified for clarity. (b) densities of various monolayers in TMD, MXene and TXene. Structural parameters for TMD and MXene are from refs. \cite{DING20112254} and \cite{kurtoglu_naguib_gogotsi_barsoum_2012}. (c) Poresizes of TXene for various compositions. Red and blue symbols are for $d_{1}$ and $d_{2}$ as marked in (a), and gray circle is the average values.}
	\label{fig1}
\end{figure}

Atomic arrangements of TXene along with typical 2D materials such as TMD and MXene are shown in Figure~\ref{fig1}(a).
Although 1H phase of TMD and M$_{2}$X of MXene are selectively shown for their structural simplicity, other phases in either TMD or MXene also show similar structural features to be discussed.
TXene shows tetrahedral bonding network which is distinct from other 2D materials containing transition metal elements which prefer closed-packing.
This can be confirmed by space groups with higher symmetry of TMD and MXene such as P$\bar{6}$m2 (No. 187) and P$\bar{3}$m1 (No. 164), respectively, while TXene is characterized as Cmme (No. 67).
In addition, thanks to strongly directional bonds of carbon group elements, open-channel structure is formed along an in-plane crystallographic orientation marked as $\hat{x}$ in Figure~\ref{fig1}(a).
Considering the unique stoichiometry and space group of TXene, we carefully examine a number of recently reported 2D crystals and their database~\cite{doi:10.1021/acs.nanolett.6b05229, PhysRevLett.118.106101, Mounet2018, 2053-1583-5-4-042002} and confirm that the present atomic arrangement of TXene is unique, and has not been reported previously elsewhere to the best of our knowledge.

The fact that TXene possesses open-framework renders a potential for energy applications for conversion and storage. 
Densities of monolayer TXene materials are significantly lower than that of TMD and MXene as in Figure~\ref{fig1}(b), enhancing higher energy and power density for a given number of active sites.
For instance, density of Sn$_{3}$Se is over two- and threefold reduced compared to that of MoSe$_{2}$ and WSe$_{2}$, respectively, of which atomic weights of the constituent elements are comparable to each other.
Furthermore, Figure~\ref{fig1}(c) indicates that those pores are large enough so that they could accommodate light metal ions such as lithium and magnesium, indicating that the pore array may serve as adsorption site or conducting channel for those ions.
Simple estimation shows that the smaller pore radius of Si$_{3}$O ($\approx$2~\AA) is greater than sum of Shannon's radius of lithium (0.76~\AA~\cite{Shannon:a12967}) and covalent radius of silicon of 1.11~\AA.
Interestingly, shape of the pores varies from elliptical to circular as size of the chalcogen elements becomes greater as in Figure~\ref{fig1}(c).

We note that bonding character of tetragen at different positions varies significantly although they are atomically thin materials.
Electrons are transferred from the surface tetragen atoms to chalcogen atoms, while the tetragen atoms still bind to each other covalently as seen in Figure~S1 in the Supporting Information (SI).
This can also be shown quantitatively by using Bader charge analysis~\cite{HENKELMAN2006354} that the bridging tetragen atoms, which bind only to other tetragen atoms, are nearly unaffected by the chalcogen atoms (Figure~S2).
On the other hand, the bonding character between the surface tetragen and chalcogen atoms is mixture of covalent and ionic bonding, and the former becomes more significant when the less electronegative chalcogen is attached.

\begin{figure*}[]
	\centering
	\includegraphics[width=\textwidth]{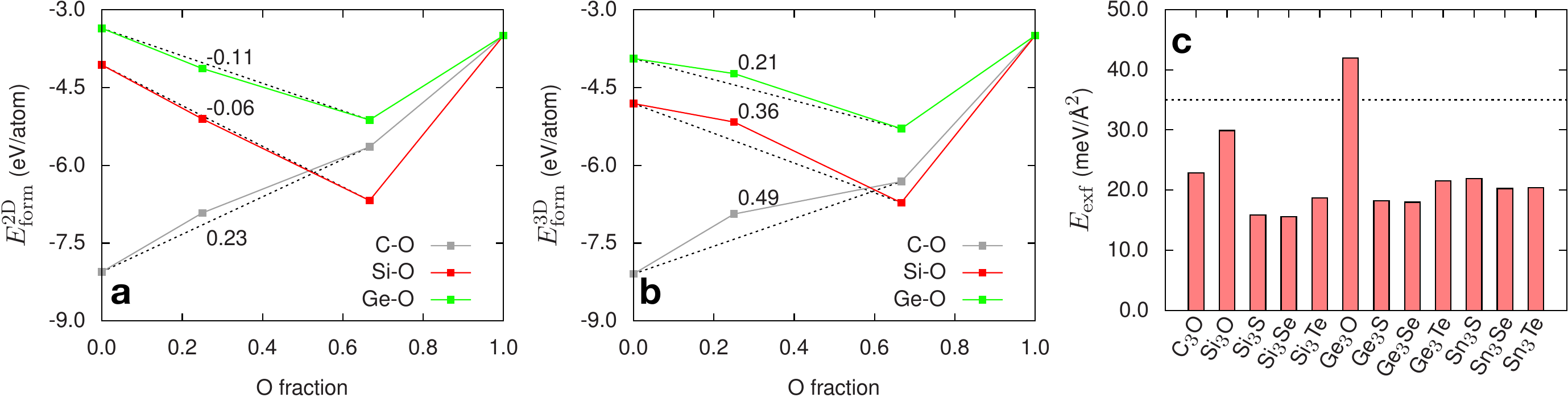}
	\caption{Stability of T$_{3}$O (T=C, Si and Ge) compounds: Convex Hull diagrams in their (a) layered bulk and (b) isolated forms. For T$_{3}$O with O fraction of 0.25, Hull distances are shown. (c) Exfoliation energy ($E_{\mathrm{exf}}$) calculated using Equation (1). The dashed line at 35 meV/\AA$^{2}$ indicates the \emph{easily exfoliable} limit suggested by Mounet \emph{et al.}~\cite{Mounet2018}. (d) Instantaneous total energy and (e) temperature for Si$_{3}$O. A snapshot at 5 ps is shown in the inset of (d).}
	\label{fig2}
\end{figure*}

Thermodynamic stability for T$_{3}$O (T=C, Si, Ge) is compared with other T-O compounds in convex Hull diagram as shown in Figure~\ref{fig2}.
Details of other T-O compounds with a varying O fraction can be found SI S2.
When compared with other bulk phases, the layered bulk form of T$_{3}$X with space group of C2/m (see Figure~S3) shows relatively high Hull distances of 0.49, 0.36 and 0.21 eV/atom for C$_{3}$O, Si$_{3}$O and Ge$_{3}$O, respectively, indicating that these phases are metastable even if they are synthesized.
However, we note that synthesizability of metastable phases involves many thermodynamic parameters~\cite{doi:10.1063/1.4944801, doi:10.1021/acs.chemmater.7b02399}, and depends critically on finding constrained equilibrium conditions~\cite{Aykoleaaq0148, Sune1600225}.
For instance, Si atoms confined within a planar pore composed of 2D CaF$_{2}$ crystals can spontaneously form new 2D silicon crystals~\cite{yaokawa_monolayer--bilayer_2016}.
Furthermore, considering recent rapid developments in growing few-layer 2D crystals directly on desired surfaces~\cite{zhou_library_2018}, relative formation energies among their isolated single layers ($E^{\mathrm{2D}}_{\mathrm{form}}$) as in Figure~\ref{fig2}(b) would be of great interest.
Interestingly, the Hull distances for T$_{3}$Xs are remarkably reduced due to the confinement.
Especially, Si$_{3}$O and Ge$_{3}$O become stable as a result of confinement with negative Hull distances of -0.06 and -0.11 eV/atom, respectively, suggesting a feasible synthesis pathway of TXenes.

Surfaces of the T$_{3}$X compounds are chemically inert sufficiently, so that most of the T$_{3}$X can form stable layered bulk phase as in Figure~S3. 
All the compounds except Ge$_{3}$O show exfoliation energies computed as in Equation (1), ranging from 15.6 to 30.0 meV/\AA$^{2}$, fall in to the \emph{easily exfoliable} category defined by Mounet \emph{et al}.~\cite{Mounet2018} as in Figure~\ref{fig2}(c). 
Even the highest exfoliation energy of 42 meV/\AA$^{2}$ for Ge$_{3}$O seems very likely to be exfoliable, which is quite close to the threshold value of 35 meV/\AA$^{2}$. 

We further demonstrate that the new T$_{3}$X compounds are dynamically stable against thermal fluctuations.
Especially for the single layer of Si$_{3}$O, we confirmed that the system remain stable at 1500 K for 10 ps in our AIMD simulations as in Figure~\ref{fig2}(d, e).
The remarkable thermal robustness ensures that the unique structural template of Cmme can provide significant energetic stability as a monolayer and energy barrier of phase transition for various T$_{3}$X compounds.
We provide harmonic phonon dispersion spectra (Figure~S5) and dynamic behaviors in MD simulations (Figure~S6) for all the other compounds in the SI.

We note that relative atomic sizes between tetragen and chalcogen play an important role to determine the stability of the crystals.
For instance, comparatively large chalcogen elements such as S, Se and Te bound to small carbon atoms make the corresponding compounds of C$_{3}$S, C$_{3}$Se and C$_{3}$Te dynamically unstable making phonon dispersion spectra have imaginary phonon modes as in Figure~S5.
On the other hand, surface of the Sn$_{3}$O becomes chemically active, so that Sn atoms in adjacent layers covalently bind to each other over the O atoms on the surface, preventing Sn$_{3}$O from forming layered structure (Figure~S4).
In addition, it is clearly seen that exfoliation energy for oxides increases monotonously when the relative size of tetragen becomes greater: i.e., from C$_{3}$O to Si$_{3}$O to Ge$_{3}$O in Figure~\ref{fig2}(c). 

\begin{figure*}[]
	\centering
	\includegraphics[width=0.9\textwidth]{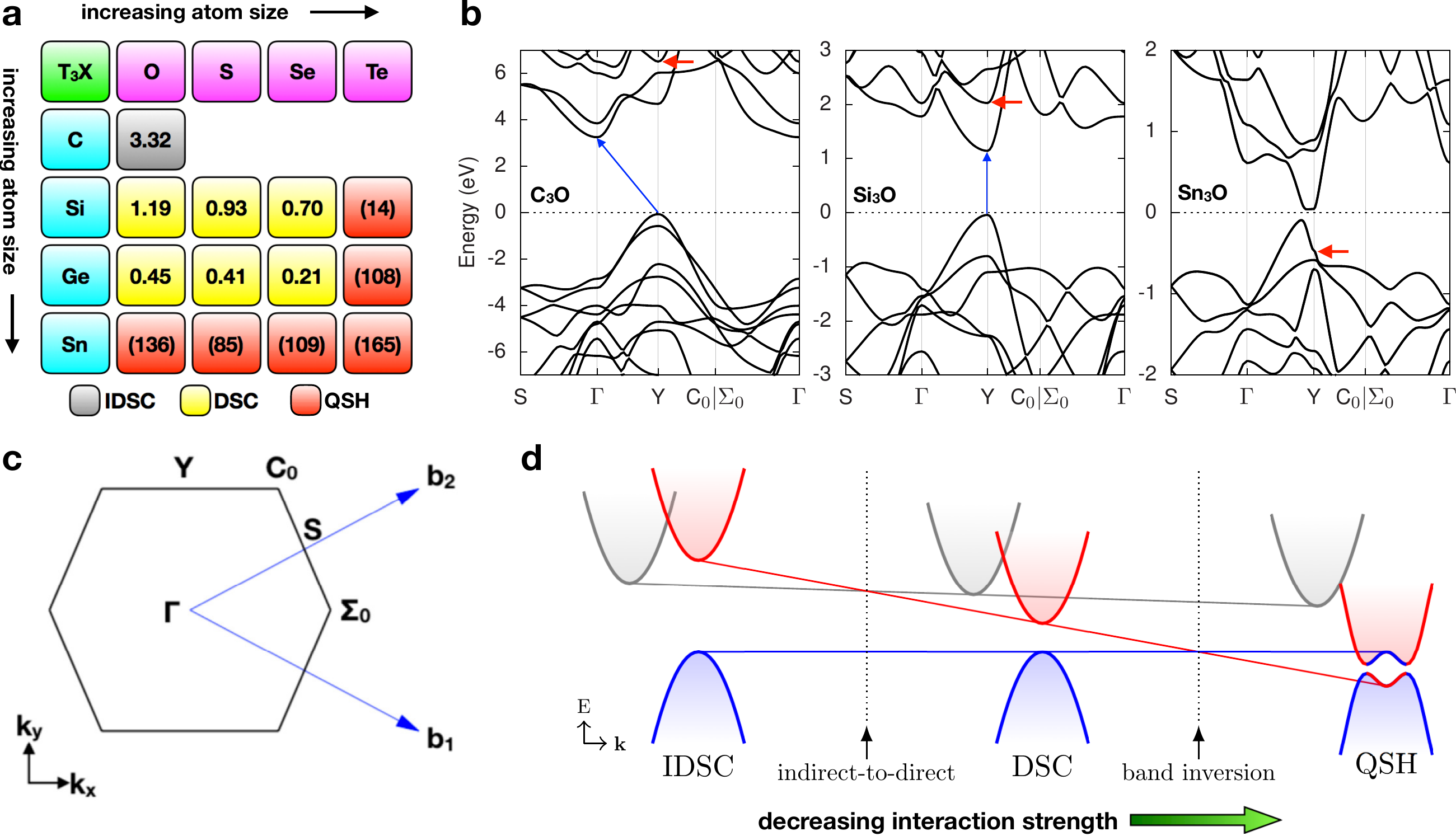}
	\caption{Electronic structures: (a) Size of band gaps for various T$_{3}$X compounds calculated with HSE06 hybrid functional. Boxes in gray and yellow shading indicate indirect and direct band gap semiconductors (IDSC and DSC), respectively, and the band gaps are given in eV. The red boxes are for quantum spin Hall insulators (QSH), and the band gap due to spin-orbit interaction is given in a parenthesis with meV unit. (b) Electronic band dispersions (HSE06) for the corresponding materials: C$_{3}$O, Si$_{3}$O and Sn$_{3}$O, respectively. Fermi level is located at 0 eV. (c) Schematic first Brillouin zone of the T$_{3}$X with high symmetry point labels. (d) Schematic diagram for electronic phase transition with respect to interaction strength, which is responsible for versatile electronic phases in TXene: IDSC, DSC and QSH.}
	\label{fig3}
\end{figure*}

Having established structural stabilities of TXenes, now we discuss their electronic characteristics.
Various T$_{3}$X compounds show versatile electronic properties.
As seen in Figure~\ref{fig3}(a), not only that they have a variety of electronic phases such as indirect or direct band gap semiconductors and topologically nontrivial quantum spin Hall insulators with spin-orbit gaps, but also that their energy gaps vary significantly for different compositions.
The computed band gap size can be as large as 3.3 eV for C$_{3}$O (indirect band gap), and is decreased down to 0.2 eV for Ge$_{3}$Se (direct band gap) for band insulators.
The spin-orbit gaps range from 14 meV for Si$_{3}$Te to 165 meV for Sn$_{3}$Te.
The highest value of 165 meV for Sn$_{3}$Te at their pristine form without mechanical strain or chemical decorations is remarkable because it is well-above the room temperature thermal fluctuations ($\sim$25 meV), and comparable to other theoretically predicted topological insulating materials with large energy gaps ranging 0.1--0.3 eV~\cite{PhysRevLett.97.236805,PhysRevX.4.011002,PhysRevLett.111.136804,doi:10.1021/acs.nanolett.5b02617}.

The band gap sizes vary differently with respect to atomic sizes, indicating different band gap opening mechanisms for band and topological insulating TXenes.
For trivial insulating TXenes, band gap becomes diminished with larger ions, but the trend is opposite for quantum spin Hall insulating ones as can be seen in Figure~\ref{fig3}(a).
This is because each of the TXene materials has the identical atomic arrangement and electron filling due to the same valence electronic configurations, but their primary bonding strengths determining the band gap size vary gradually with their sizes; the smaller the sizes, the stronger the interactions. 
Versatile electronic phases and sizable band gaps in various TMDs were explained similarly~\cite{Chhowalla2013}.
We confirm this from the calculated electronic band dispersions as in Figure~\ref{fig3}(b); the bands marked as red arrows in Figure~\ref{fig3}(b) have quite similar wave function characters, while their relative energetic positions vary with chemical compositions.
Note that energetic position of this band plays a critical role, so that we indicate it as $\varepsilon^{*}_{Y}$ for further discussion.
The band is mainly composed of p$_{y}$ and p$_{z}$ orbital of the surface atoms, and shows antibonding character.
Thus, $\varepsilon^{*}_{Y}$ goes down with decreasing interactions or increasing size.

Figure~\ref{fig3}(d) shows schematic depictions accounting for various electronic structures of TXene.
To begin with C$_{3}$O with the strongest interactions, the $\varepsilon^{*}_{Y}$ is located higher than other bands in the Brillouin zone in Figure~\ref{fig3}(b), making it an indirect band gap semiconductor.
On the other hand, bonding in Si$_{3}$O is weaker than that of C$_{3}$O and the $\varepsilon^{*}_{Y}$ is shifted down to make the system a direct band gap semiconductor.
Note that $\varepsilon^{*}_{Y}$ seems to vary the most sensitively to the interaction strength compared to the conduction bands with different momenta.
Further decrease of interaction strength make the $\varepsilon^{*}_{Y}$ continue to be decreased, and eventually make the two band edges near the Fermi level inverted with respect to each other, realizing quantum spin Hall insulator; see Figure~\ref{fig3}(b) and (d).
Thanks to presence of inversion symmetry, we confirmed the Z$_{2}$ topological invariant of the systems with band inversion to be 1, identifying those inverted bands topologically nontrivial.

It is worth noting here that prediction of a robust structural template that can host various chemical compositions will help accelerate discovery of novel materials in many ways.
Firstly, a series of novel materials sharing the same atomic arrangement but with different interaction strength will display a variety of properties, e.g., electronic properties in Figure~\ref{fig3}(a), thus extending property spectra by exploring material space.
Furthermore, the predicted structural prototype can also be used to further find novel materials by performing high-throughput calculations, which can expand elements for substitution to a variety of elements in the periodic table. 
The 2D CSP method used in this work (SANDWICH~\cite{2053-1583-5-2-025013}) is highly efficient and transferable, and is anticipated to be used to uncover other 2D structural prototypes which have not been explored yet.
Lastly, the rich variety of materials database containing novel materials with wide ranges of structure-property relations can be used for predictive models based on machine learning and artificial intelligence.

\begin{figure}[]
	\centering
	\includegraphics[width=\columnwidth]{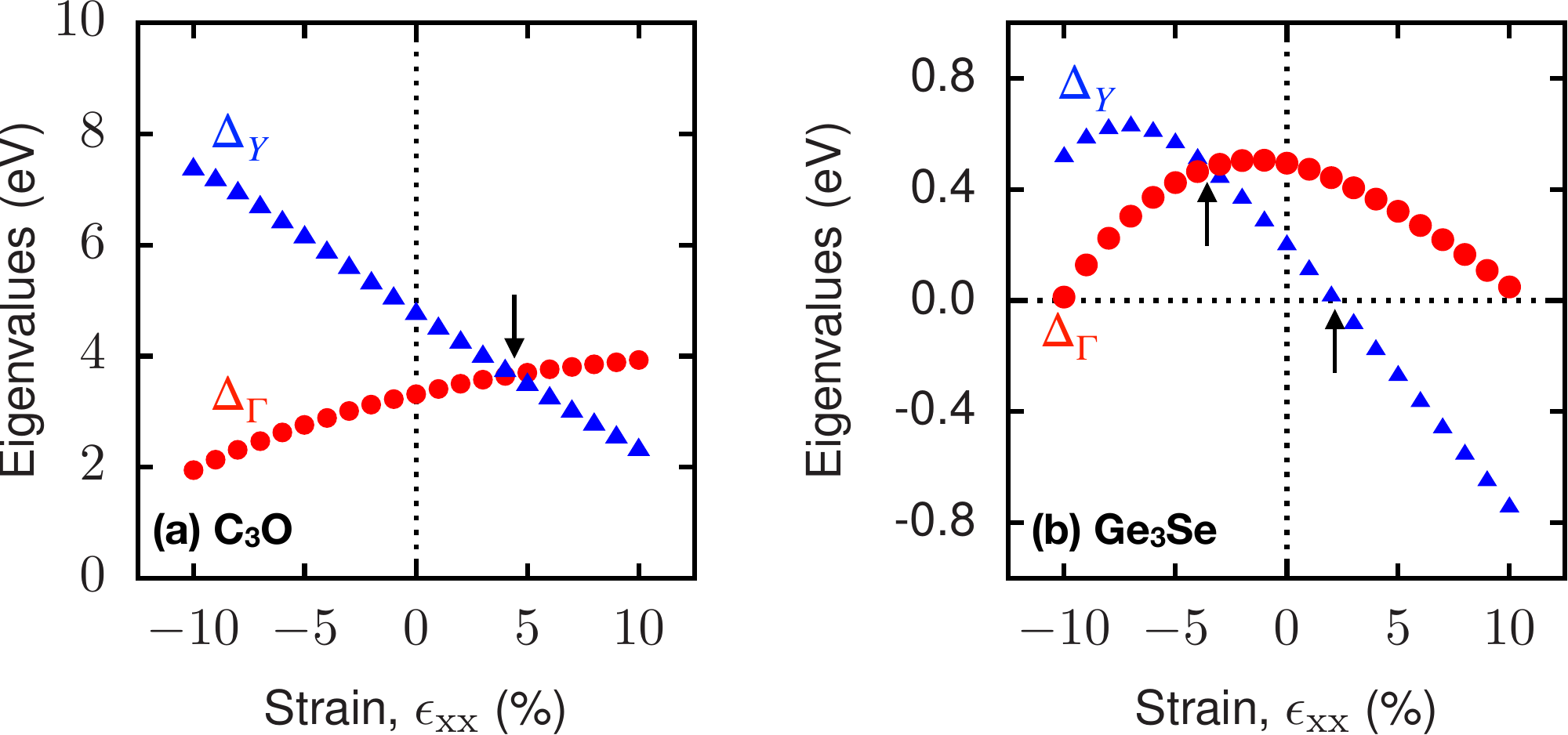}
	\caption{Effects of uniaxial strain along x ($\epsilon_{\mathrm{xx}}$). (a) Indirect-to-direct transition of C$_{3}$O and (b) trivial-to-nontrivial transition of Ge$_{3}$Se. The critical strains at which transitions occur are marked as arrows. $\Delta_{\Gamma}$ and $\Delta_{Y}$ are energetic distances of the unoccupied band from the valence band maximum.}
	\label{fig4}
\end{figure}

The diverse electronic properties, previously shown to be controlled primarily by bonding strength of the surface atoms, can be tuned further by mechanical strain.
In particular, we show the effects of uniaxial strain along $\hat{x}$ ($\epsilon_{\mathrm{xx}}$) which is along the zigzag chain of tetragen atoms on the surface as seen in Figure~\ref{fig1}(a).
It is confirmed that not only energetic gaps change quantitatively according to the uniform mechanical strain on the system, electronic phases of individual T$_{3}$X systems may also be altered qualitatively as well.
As discussed previously, the $\varepsilon^{*}_{Y}$ plays a crucial role to determine the electronic property.
Therefore, we examine the relative positions of $\varepsilon^{*}_{Y}$ and $\varepsilon^{*}_{\Gamma}$ (the empty band at momentum of $\Gamma$) with respect to the valence band maximum. 
For instance, $\Delta_{Y}$ is defined as energy difference between $\varepsilon^{*}_{Y}$ and the valence band maximum.
In all the cases, the $\Delta_{Y}$ decreases with positively increasing $\epsilon_{\mathrm{xx}}$, while the $\Delta_{\Gamma}$ responds oppositely as in Figure~\ref{fig4}.
Thus, for positive uniaxial strain along the $\hat{x}$-axis, there is a crossover between $\varepsilon^{*}_{Y}$ and $\varepsilon^{*}_{\Gamma}$ at $\epsilon_{\mathrm{xx}}$ of $\sim$4 \% (marked as arrow) showing indirect-to-direct band gap transition for C$_{3}$O.
Similarly, Si$_{3}$Se, a small direct band gap semiconductor, becomes a quantum spin Hall insulator when it is slightly elongated along the surface chains, while it remains a direct band gap semiconductor otherwise.
The negative $\epsilon_{\mathrm{xx}}$ at around -3.5 makes the system have indirect band gap in Figure~\ref{fig4}(b).

\section{Conclusion}
In conclusion, we demonstrate theoretical prediction of a structural prototype for a new family of 2D crystals, T$_{3}$X, or TXene, composed of group IV (T) and group VI (X) elements with tetrahedral bonding network as well as open-framework.
We confirmed 13 stable monolayers of 2D T$_{3}$X compounds, and most of them also formed stable layered bulk phase with their binding energies falling in desired range.
In particular, a significant thermal stability was confirmed for Si$_{3}$O monolayer by $\emph{ab initio}$ molecular dynamics simulations at elevated temperature.
Moreover, the T$_{3}$X compounds show versatile electronic structures, which is mainly due to the varying interaction strengths, which are tunable by mechanical strain.
The novel 2D structural prototype with a number of stable compounds already found and more to be found potentially may not only demonstrate intriguing material properties, but also accelerate novel materials discovery.

\begin{acknowledgements}
We thank Korea Institute for Advanced Study for providing computing resources (KIAS Center for Advanced Computation Linux Cluster System) for this work. Y.-W.S. was supported by NRF of Korea (Grant No. 2017R1A5A1014862, SRC program: vdWMRC Center).
\end{acknowledgements}

%

\clearpage
\widetext

\begin{center}
\textbf{\large Supplementary Material for: \\A New Family of Two-Dimensional Crystals: Open-Framework T$_{3}$X (T=C, Si, Ge, Sn; X=O, S, Se, Te) Compounds with Tetrahedral Bonding} \\
\vspace{10pt}
Kisung Chae and Young-Woo Son$^{*}$ \\
\vspace{4pt}
\emph{Korea Institute for Advanced Study, Seoul 02455, South Korea}\\
(Dated: \today)
\end{center}

\renewcommand{\thesection}{S\Roman{section}}
\setcounter{section}{0}
\renewcommand{\thefigure}{S\arabic{figure}}
\setcounter{figure}{0}
\renewcommand{\thetable}{S\arabic{table}}
\setcounter{table}{0}
\renewcommand{\bibnumfmt}[1]{[S#1]}
\renewcommand{\citenumfont}[1]{S#1}

\section{S1. Bonding characters of T$_{3}$X compounds}
\begin{figure}[h]
	\centering
	\includegraphics[width=0.6\textwidth]{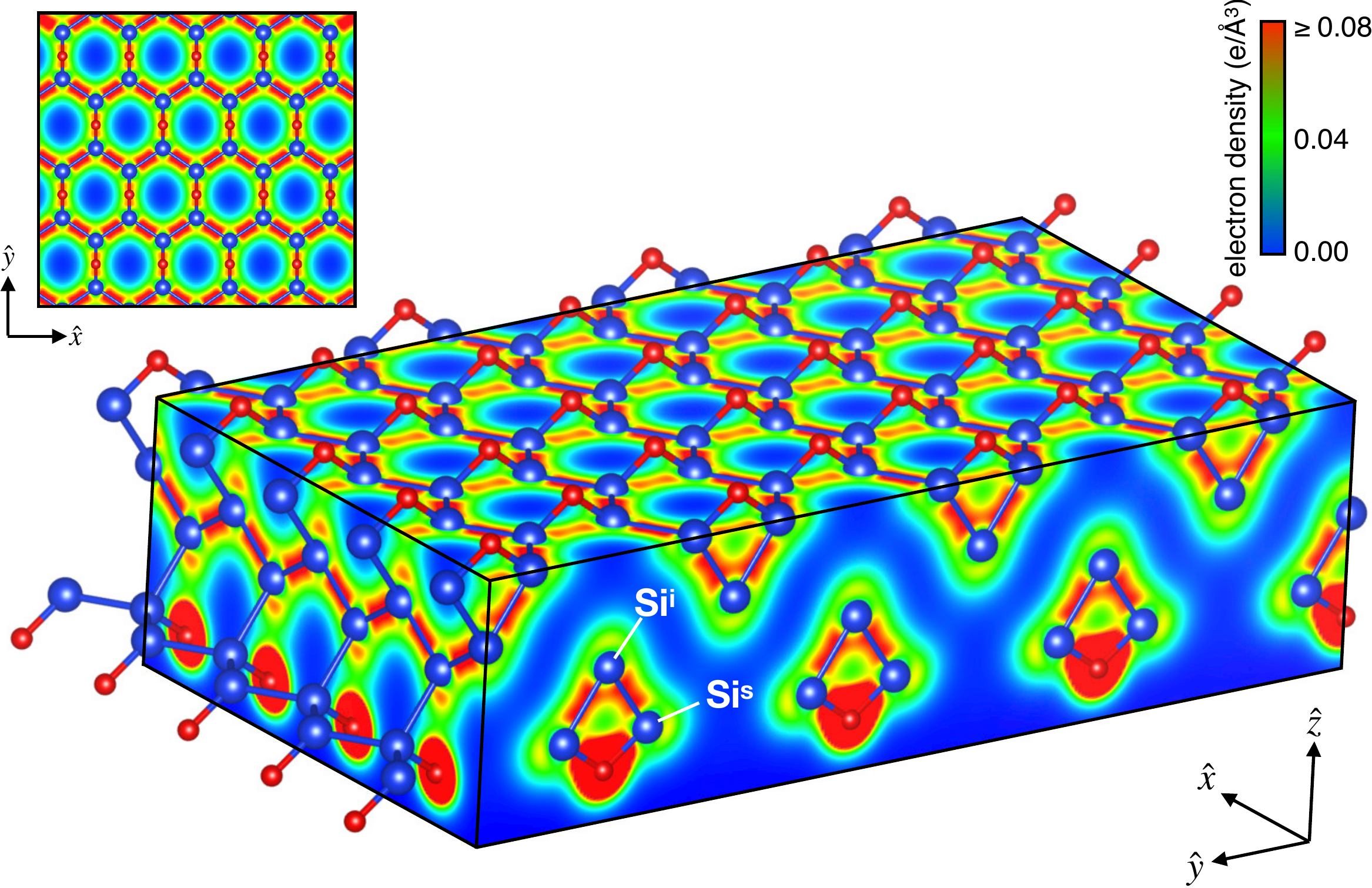}\\
	\caption{Charge density plot of Si$_{3}$O. The xy-plane projection is shown in the inset for clarity. The different types of Si atoms at different locations are marked as Si$^{s}$ and Si$^{i}$.}
	\label{figs_chgden}
\end{figure}

\begin{figure}[h]
	\centering
	\includegraphics[width=0.5\textwidth]{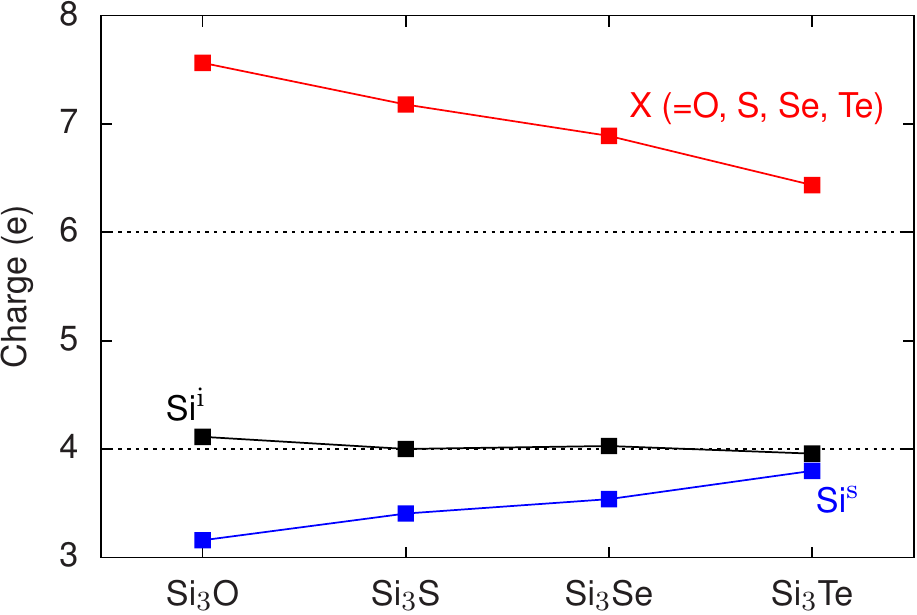}\\
	\caption{Charge of each atom by Bader partitioning for silicon compounds. Si$^{s}$ and Si$^{i}$ refer to silicon atoms located at surface and inside, respectively.}
	\label{figs_bader}
\end{figure}

To better understand the bonding characters, we show the charge density for Si$_{3}$O in Figure~\ref{figs_chgden}. We also show Bader charges for silicon compounds in Figure~\ref{figs_bader}.

\section{S2. Phases used for convex Hull diagram}
To discuss Thermodynamic stability of T$_{3}$X compounds, we generated convex Hull diagram as in Figures~2(a, b).
Various T-O phases (T=C, Si, Ge) with a varying chemical composition are listed in the Table~\ref{tab1}.
The three-dimensional (3D) structures were found from Open Quantum Materials Database (OQMD)~\cite{Saal2013, kirklin_open_2015}.
For two-dimensional (2D) structures, we researched some literature data, and found 2D structures of SiO$_{2}$~\cite{PhysRevLett.95.076103} and GeO$_{2}$~\cite{PhysRevB.95.155426}.
For 2D CO$_{2}$, we used the same structure as SiO$_{2}$.

\begin{table}[htp]
\caption{Various T-O phases (T=C, Si, Ge) used to generate convex Hull diagram in Figure~2(a, b).}
\begin{center}
\begin{tabular}{lllll}
\hline
\hline
dimension & composition & O fraction & space group & reference \\
\hline
\multirow{12}{*}{3D} & C & 0.00 & P6/mmm & - \\
 & CO$_{2}$ & 0.67 & Pa$\bar{3}$ & \cite{Simon:a19264} \\
 & C$_{3}$O & 0.25 &  C2/m & this work \\
 & O & 1.00 & C2/m & \cite{PhysRevLett.97.085503} \\
\cline{2-5}
 & Si & 0.00 & Fd$\bar{3}$m & - \\
 & SiO$_{2}$ & 0.67 & I$\bar{4}$2d & \cite{PhysRevB.78.054117} \\
 & Si$_{3}$O & 0.25 & C2/m & this work \\
 & O & 1.00 & C2/m & \cite{PhysRevLett.97.085503} \\
\cline{2-5}
 & Ge & 0.00 & Fd$\bar{3}$m & - \\
 & GeO$_{2}$ & 0.67 & P4$_{2}$/mnm & \cite{Baur:a08385} \\
 & Ge$_{3}$O & 0.25 & C2/m & this work \\
 & O & 1.00 & C2/m & \cite{PhysRevLett.97.085503} \\
\hline
\hline
\multirow{12}{*}{2D} & C & 0.00 & P6/mmm & - \\
 & CO$_{2}$ & 0.67 & P6/mmm & \cite{PhysRevLett.95.076103} \\
 & C$_{3}$O & 0.25 & Cmme & this work \\
 & O & 1.00 & C2/m & \cite{PhysRevLett.97.085503} \\
\cline{2-5}
 & Si & 0.00 & P$\bar{3}$m1 & - \\
 & SiO$_{2}$ & 0.67 & P6/mmm & \cite{PhysRevLett.95.076103} \\
 & Sn$_{3}$O & 0.25 & Cmme & this work \\
 & O & 1.00 & C2/m & \cite{PhysRevLett.97.085503} \\
\cline{2-5}
 & Ge & 0.00 & P$\bar{3}$m1 & - \\
 & GeO$_{2}$ & 0.67 & C2/m & \cite{PhysRevB.95.155426} \\
 & Ge$_{3}$O & 0.25 & Cmme & this work \\
 & O & 1.00 & C2/m & \cite{PhysRevLett.97.085503} \\
\hline
\hline
\end{tabular}
\end{center}
\label{tab1}
\end{table}

\newpage
\section{S3. Bulk phases of T$_{3}$X compounds}
\begin{figure}[h]
	\centering
	\includegraphics[width=0.75\textwidth]{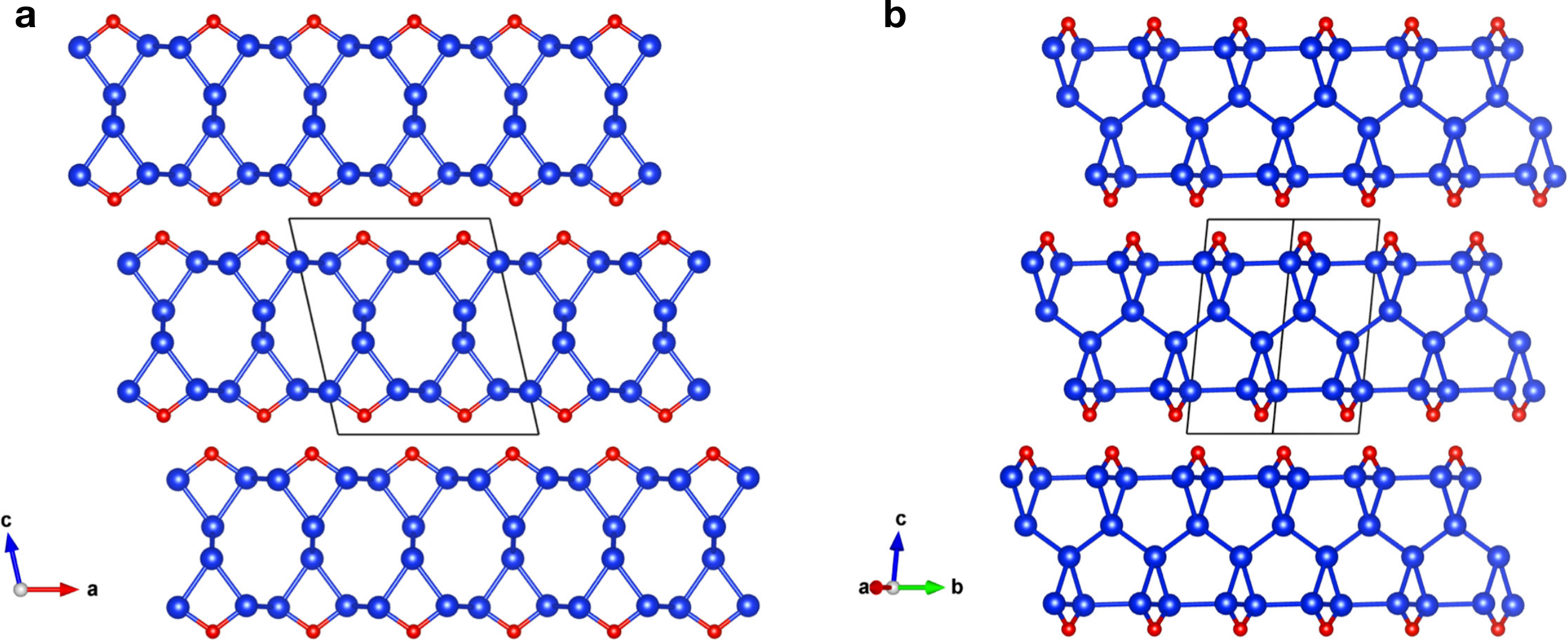}\\
	\caption{Layered structure of TXene in a monoclinic phase projected along (a) b axis and (b) face diagonal orientations.}
	\label{figs_bulk}
\end{figure}

\begin{figure}[h]
	\centering
	\includegraphics[width=0.75\textwidth]{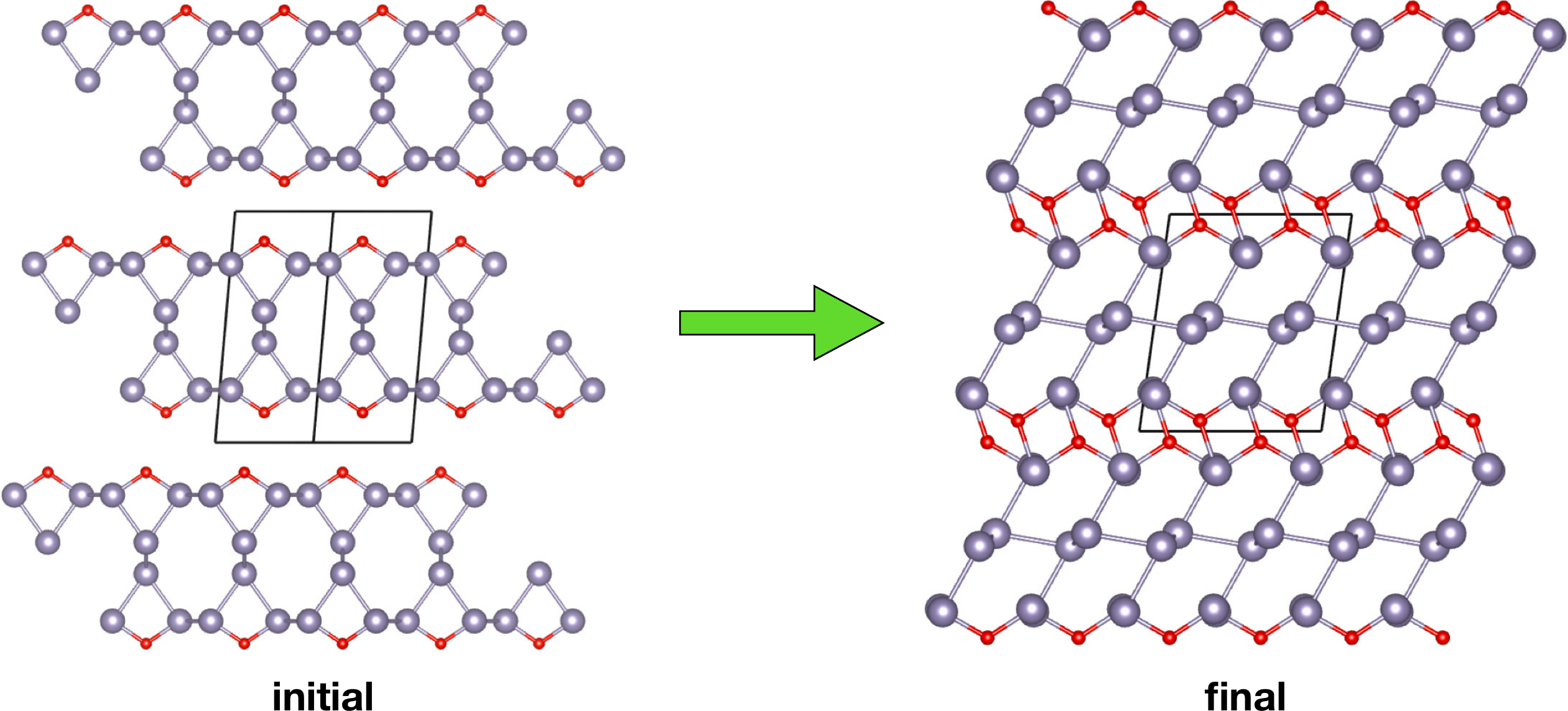}\\
	\caption{Initial and final structures of structural optimization for Sn$_{3}$O.}
	\label{figs_sn3o_bulk}
\end{figure}

Most of the stable T$_{3}$X compounds can form layered structures with space group of C2/m as in Figure~\ref{figs_bulk}, except Sn$_{3}$O of which the atomic size ratio between tetragen and chalcogen is the greatest.
As we discussed in the main text, this makes the surface of the Sn$_{3}$O chemically reactive, so that Sn atoms in adjacent layers chemically bind to each other, resulting in the bulk Sn$_{3}$O not layered structure any more; see Figure~\ref{figs_sn3o_bulk}.

\newpage
\section{S4. Dynamical stability for all the T$_{3}$X compounds}
\begin{figure}[h]
	\centering
	\includegraphics[width=0.5\textwidth]{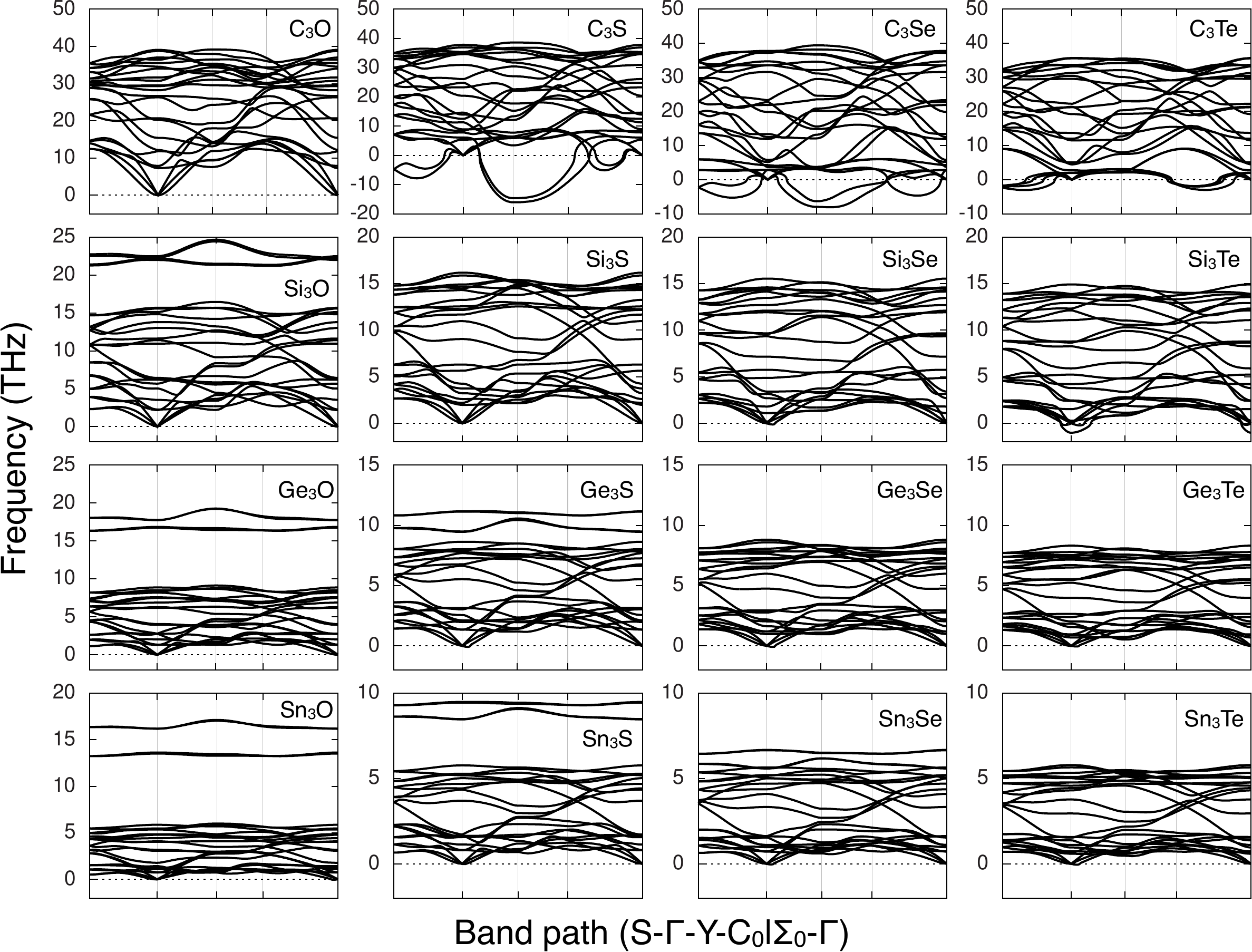}\\
	\caption{Phonon dispersion spectra for all the T$_{3}$X compounds. Same band path was used as electronic band dispersion in Figure 2.}
	\label{figs_phonon}
\end{figure}

We investigated dynamical stability of the T$_{3}$X compounds by harmonic phonon dispersion relations as in Figure~\ref{figs_phonon}.
Except some of the carbon-based compounds such as C$_{3}$S, C$_{3}$Se and C$_{3}$Te, total of thirteen T$_{3}$X compounds out of the sixteen show stable harmonic phonon dispersion.
It is interesting that the Sn$_{3}$O monolayer is shown to be dynamically stable, while its bulk is not as discussed above (Figure~\ref{figs_sn3o_bulk}).
Thus, monolayer Sn$_{3}$O may be still useful when they are encapsulated properly.

%

\end{document}